 \documentclass[final,letterpaper,twoside,12pt]{article}

\usepackage{epsfig,graphicx,psfrag}
\usepackage{epsfig,graphicx,psfrag,bm,amssymb, setspace}

\usepackage{graphicx}
\usepackage{dcolumn}
\usepackage{hepunits}
\usepackage{amsmath,amssymb,amsfonts}
\usepackage{mathrsfs}
\usepackage{color}
\usepackage{hyperref}
\usepackage{subfigure}

\newcommand{\be}{\begin{eqnarray}}
\newcommand{\ee}{\end{eqnarray}}

\begin{document}
\baselineskip 0.6cm

\def\simgt{\mathrel{\lower2.5pt\vbox{\lineskip=0pt\baselineskip=0pt
           \hbox{$>$}\hbox{$\sim$}}}}
\def\simlt{\mathrel{\lower2.5pt\vbox{\lineskip=0pt\baselineskip=0pt
           \hbox{$<$}\hbox{$\sim$}}}}
\def\one{\relax{\rm 1\kern-.25em 1}}

\def\lsim{\mathrel{\rlap{\lower4pt\hbox{\hskip0.2pt$\sim$}}
 \raise2pt\hbox{$<$}}}

\def\gsim{\mathrel{\rlap{\lower4pt\hbox{\hskip0.2pt$\sim$}}
 \raise2pt\hbox{$>$}}}

\begin{titlepage}

\begin{flushright}
\end{flushright}

\vskip 1.0cm

\begin{center}

{\Large \bf 
Reach in All Hadronic Stop Decays: \\ \vskip 0.3cm
A Snowmass White Paper
}

\vskip 0.6cm

{\large
Daniel Stolarski\footnote{\tt danchus@umd.edu}
}

\vskip 0.4cm

{\it Department of Physics and Astronomy, Johns Hopkins University, 
         Baltimore, MD 21218} \\
{\it Center for Fundamental Physics, Department of Physics, \\
         University of Maryland, College Park, MD 20742} \\

\vskip 0.8cm

\abstract{\noindent We study the discovery prospects for stops which decay to a top and a light neutralino. We consider fully hadronic decays of the tops and present an estimate for the reach at various future collider runs. Our results are summarized in Table~\ref{tab:results}.
}

\end{center}
\end{titlepage}

\setcounter{tocdepth}{2}
\singlespacing
\tableofcontents
\singlespacing

\section{Introduction}

Supersymmetry has long been considered a leading solution to the hierarchy problem~\cite{Martin:1997ns}. The largest radiative correction to the Higgs potential arises from top loops, thus the scalar partner of the top (stop) is of critical importance for understanding if supersymmetry solves the hierarchy problem. Therefore, in this work, we will study the reach for stops at future high energy hadron colliders. Motivated by dark matter and proton decay, we consider $R$-parity to be a good symmetry and imagine a neutral lightest supersymmetric particle (LSP) that is  stable on collider time scales. We will refer to the LSP as a neutralino ($\chi^0$), but it could have quantum numbers which differ from the usual MSSM neutralinos. Thus we take a simplified model which consists of a stop and a much lighter neutralino, and, in this model, the decay $\tilde{t}\rightarrow t \chi^0$ occurs 100\% of the time. 

Other Snowmass reports and many papers in the literature consider searching for stops using events with leptons. Here we consider fully hadronic decays using strategies inspired by~\cite{Plehn:2010st,Plehn:2012pr,Kaplan:2012gd,Dutta:2012kx,Buckley:2013lpa}. Current experimental searches for this channel include~\cite{ATLAS-CONF-2013-024,:2012si,CMS-PAS-SUS-11-030,CMS-PAS-SUS-11-024}, with~\cite{ATLAS-CONF-2013-024} placing the best limit on this simplified model. The fully hadronic channel has two advantages over leptonic searches. The first is that it has the largest branching fraction for the top decays. The second is that it has no inherent missing energy from neutrinos, so all the missing energy comes from the neutralinos. This allows many backgrounds to be reduced by vetoing events with leptons. 

Here we will present a very crude estimate of the reach at various colliders. We will choose stringent cuts to get a very pure signal sample, and then compute the signal efficiency using literature and simplified parton level simulations. We will then convert that efficiency into a reach using tree level cross sections given by MadGraph 5~\cite{Alwall:2011uj}. The results are summarized in Table~\ref{tab:results}.

\begin{table}[t]
\centering
\begin{tabular}{|c|c|c|c|c|}
\hline
	Collider	&	Energy		&	Luminosity		&		Cross Section &  Mass   \\       \hline		\hline
	LHC8 	&	8 TeV		&	20.5 fb$^{-1}$			&  10 fb     &	650 GeV                \\ \hline
	LHC 	&	14 TeV			&	300 fb$^{-1}$			&  3.5 fb     &	1.0 GeV                \\
	HL LHC	&	14 TeV		&	3 ab$^{-1}$			&  1.1 fb   &	1.2 TeV    	\\
	HE LHC &    	33 TeV                         &       3 ab$^{-1}$                &   91 ab   &      3.0 TeV     \\	
	VLHC &    	100 TeV                         &       1 ab$^{-1}$                     &    200 ab     &   5.7 TeV     \\	
\hline
\end{tabular} \hspace{-0.138cm}\vline
\vspace{0.3cm}
\caption{The first line gives the current bound on stops from the LHC~\cite{ATLAS-CONF-2013-024}. The remaining lines give the estimated 5$\sigma$ discovery reach in stop pair production cross section and mass for different future hadron collider runs.  }
\label{tab:results}
\end{table}%

\section{Search Strategy}

The fully hadronic channel is given by $\tilde{t}\tilde{t}^*\rightarrow t\bar{t} + \chi^0 \chi^0$ with a final state of $bjjbjj\chi^0\chi^0$. We will try to design cuts to get a very pure signal sample.

\subsection{Signal Efficiency}
\label{sec:signal}

We use top tagging~\cite{Thaler:2008ju, Kaplan:2008ie,Almeida:2008yp,Plehn:2010st,Thaler:2011gf,Soper:2012pb,Schaetzel:2013vka} to distinguish signal from background. For more Snowmass studies on top quark reconstruction see~\cite{Calkins:2013ega}. The general idea is to look for fat jets which exhibit substructure which is more like a hadronic top quark than various backgrounds. This is applied to stop searches in~\cite{Plehn:2010st,Plehn:2012pr,Kaplan:2012gd,Dutta:2012kx,Buckley:2013lpa}. Top tagging has been used at both CMS~\cite{Chatrchyan:2012ku} and ATLAS~\cite{Aad:2012raa} in other types of searches, and from the CMS search we take the efficiency of top tagging to be 50\% for tops with $p_T > 500$ GeV. From the same search we take the fake rate to be 5\% for the same $p_T$ range. There is very little data for $p_T > 800$ GeV, but we will use these efficiencies throughout out study, even at very high energy. The HPTTopTagger~\cite{Schaetzel:2013vka} study focuses on $p_T > 1$ TeV and finds lower tagging efficiency but also lower fake rates. 

\begin{figure}
\centerline{\includegraphics[width=.6\textwidth]{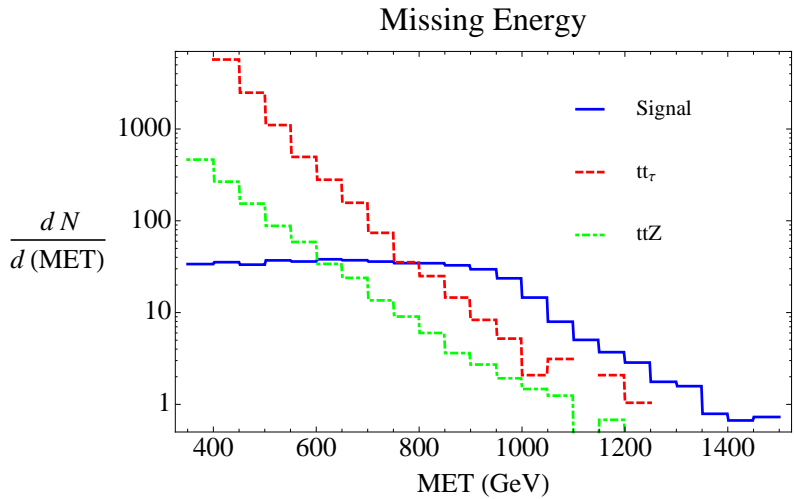}}
\caption{Distribution of missing transverse energy for the signal and the dominant backgrounds at LHC14. Event numbers are normalized to 300 fb$^{-1}$. The solid (blue) line is a 1 TeV stop signal described in Section~\ref{sec:signal} with both tops decaying hadronically, the dashed (red) line is semi-leptonic top discussed in Section~\ref{sec:tt}, and the dot-dashed (green) line is $t\bar t Z$ background discussed in Section~\ref{sec:ttz} with both the tops decaying hadronically and the $Z$ decaying invisibly. No other cuts are applied here. }
\label{fig:metstop}
\end{figure}

Therefore, we make the following cuts taking the efficiency from the literature:
\begin{itemize}
\item Require both tops decay hadronically (46\%)
\item Require one $b$-tag (70\%)~\cite{CMS-PAS-BTV-11-001,ATLAS-CONF-2011-102}
\item Require both tops pass a top tagger (25\%).
\end{itemize}
We also simulate pair production of 1 TeV stops decaying to a nearly massless (1 GeV) neutralino. The simulation is done at parton level with MadGraph 5~\cite{Alwall:2011uj} and is used to compute the efficiency for the following two cuts:
\begin{itemize}
\item Require that both tops have $p_T > 500$ GeV (19\%)
\item Require missing transverse energy bigger than 600 GeV (34\%).
\end{itemize}
The first cut justifies the efficiency of the top tagger cut from above. The efficiency of the second cut is computed after the first cut is applied. In Figure~\ref{fig:metstop} we show the missing energy distribution of the signal and dominant backgrounds, and we see the large missing energy cut is very effective in distinguishing the rapidly falling backgrounds from the relatively flat signal. 

From these cuts we estimate the total signal efficiency for stops with mass around 1 TeV to be 0.52\%. Moving in the $m_{\tilde t}-m_\chi$ plane will change the efficiency of these cuts, but our results will be applicable for stop masses around 1 TeV with very light neutralinos. 

For the High Luminosity LHC run, there will be significant pileup in each event with an expected average of 140 interactions per crossing. This could cause significant difficulties for both of our major signal discriminators, missing transverse energy and top tagging. In~\cite{Avetisyan:2013yna}, it was shown that in $t\bar t$ plus vector production, the missing energy distributions above 200 GeV are relatively stable under pileup. Since that signal topology is quite similar to those considered here, we will continue to use our parton level distributions as a crude estimate. Similarly,~\cite{Calkins:2013ega} shows that jet subtraction~\cite{Cacciari:2007fd} or grooming~\cite{Butterworth:2008iy,Ellis:2009me,Krohn:2009th} methods can make top reconstruction at high $p_T$ is relatively insensitive to pileup. Therefore, we use the same efficiencies for the cuts at high luminosity and take the same total signal efficiency of 0.52\% for the HL LHC.

For a 33 TeV collider, there are fewer studies and it becomes more difficult to extrapolate efficiencies. Nevertheless, we proceed with a crude estimate by taking the same $b$ and top tagging efficiencies and fake rates as our LHC estimate. We also continue to require that the both tops have $p_T > 500$ GeV. Because the stops now have much more energy, some of which is imparted into the neutralino, we now require missing transverse energy greater than 2.0 TeV. We simulate pair production of 3 TeV stops decaying to nearly massless neutralinos, and estimate the efficiency of the combination of the top $p_T$ cut with the missing energy cut to be 35\%, for an overall signal efficiency at 33 TeV of 2.8\%. 

For 100 TeV, all the caveats of the previous paragraphs apply but even more strongly. We continue to take tops with  $p_T > 500$ GeV and all other efficiencies the same, but we now require missing energy greater than 4.0 TeV. We take our benchmark stop mass to be 6 TeV, and derive Monte Carlo efficiency on the top $p_T$ and missing energy cuts of 37\%, for an overall efficiency of 3.0\%. A summary of the missing energy cuts and signal efficiencies used for the different colliders is given in Table~\ref{tab:bg}.

\begin{table}[t]
\centering
\begin{tabular}{|c|c|c|c|c|}
\hline
	Collider	& Missing Energy & $\varepsilon$(Signal)	& $\sigma(t\bar t _\tau) $		&	$\sigma(t\bar tZ) $		\\      \hline		\hline
	LHC  	& 600 GeV & 0.52\% 	& 1.7 ab			&	2.2 ab			                \\
	HL LHC	& 600 GeV & 0.52\% 	&1.7 ab 		&	2.2 ab			    	\\
	HE LHC & 2.0 TeV &   2.8\% 	&  $<0.1$ ab                         &      0.78 ab                 \\	
	VLHC &  4.0 TeV  & 3.0\% &  	$<0.1$ ab                        &       1.4 ab                       \\	
\hline
\end{tabular} \hspace{-0.138cm}\vline
\vspace{0.3cm}
\caption{A description of the analysis at different colliders. The parameters of the different colliders can be found in Table~\ref{tab:results}. The second column gives the missing energy cut, and the third column is the signal efficiency for all the cuts described in Section~\ref{sec:signal}.  The fourth and fifth columns are the effective cross sections after all cuts for the two dominant backgrounds: $t\bar t$ production with one top decaying to a $\tau$ and $t\bar t Z$ production.}
\label{tab:bg}
\end{table}%

\subsection{Dominant Backgrounds}
\label{sec:bg}

In order to estimate the size of the backgrounds, we use the same combination of cut efficiencies obtained from the literature and parton level Monte Carlo. Because of our requirement of  $b$ and top tags, the dominant backgrounds will be those with on-shell tops. As described below, the dominant backgrounds are semi-leptonic top production where one top decays to a $\tau$, and $t \bar tZ$ production with the $Z$ decaying invisibly. Other processes with on shell tops such as four top production and $t \bar tZZ$ have cross sections which are too small to contribute, but there may be more exotic backgrounds which contribute at super LHC energies which are not accounted for here.

\subsubsection{Semi-leptonic $t\bar t$}
\label{sec:tt}

In current 7 and 8 TeV searches, the dominant background is from $t\bar t$ production where one of the tops decays leptonically.  In order to pass the hard missing energy, the top that decays leptonically imparts substantial momentum to the neutrino in that decay. Furthermore, a veto on events with isolated leptons can eliminate most of the leptonic top decays to electron and muons. Thus we take one of our dominant backgrounds to be
$t\bar t$ production where one of the tops decays to a $\tau$: $pp \rightarrow t\bar t \rightarrow b\tau\nu_\tau bjj$, where $j$ is a light flavor jet.

The tree level cross section for this background at the LHC is 51 pb, orders of magnitude larger than the signal cross sections we are considering.  We require that the background event has a neutrino from the $W$ decay to have at least 600 GeV $p_T$. These events will also have a neutrino from the $\tau$ decay, but that one is typically much softer, so we ignore it. In Figure~\ref{fig:metstop} we see that the missing energy computed in this way falls very rapidly so that a large cut can reduce the cross section significantly. 

We also require that the top which decayed hadronically have a $p_T$ greater than 500 GeV. This, along with the missing energy cut, gets the background cross section down to 1.0 fb. For an event with one real hadronic top to pass a double top tag, it must have hard jet radiation in addition to the top quarks, and we estimate that to reduce the cross section further down to 0.1 fb. Finally, we apply the cuts that were applied to the signal:
\begin{itemize}
\item Require one $b$-tag (70\%)
\item Require the real top pass the top tagger (50\%)
\item Require the fake top pass the top tagger (5\%).
\end{itemize}
This reduces the effective background cross section to 1.7 ab at the LHC. 

At 33 TeV, the total cross section for $t\bar t$ production with one $\tau$ decay is 307 pb. Requiring the hadronic top to have $p_T > 500$ GeV, and the $\tau$ neutrino to have $p_T > 2.0$ TeV, and one hard jet, gets the cross section down to 1.5 ab, with the enormous missing energy cut doing most of the heavy lifting. Applying the $b$ and top tagger cuts, this reduces the background cross section to below 0.1 ab, making it negligible at 33 TeV, and even less important at 100 TeV with our 4 TeV missing energy cut. 

\subsubsection{$t \bar tZ$}
\label{sec:ttz}

With very hard missing energy cuts the other dominant background is $t\bar t Z$ where the $Z$ decays to neutrinos. The $t\bar t Z$ process has the same effective final states as the signal, two tops and missing energy, but the kinematic distributions will look very different. At 8 TeV, the cross section for this process is small and it is unimportant, but it gets relatively larger as we go to higher energy colliders and becomes the most important background at super-LHC energies.

The production cross section at the LHC for $t\bar tZ$ is 660 fb, so it has a much smaller raw cross section than the semi-leptonic top background, but it has two real tops, so it looks more signal like. At the LHC, the typical $p_T$ of both the tops and the $Z$ is of order the top mass, so requiring that the missing energy is bigger than 600 GeV and that both tops have $p_T> 500$ GeV reduces the cross section to 140 ab. The effect of the missing energy cut is shown in Figure~\ref{fig:metstop}. We further require:
\begin{itemize}
\item $Z$ decays to invisible (20\%)
\item Both tops decay hadronically (46\%)
\item One $b$-tag (70\%)
\item Both tops pass the top tagger (25\%).
\end{itemize}
This reduces the effective cross section for $t\bar tZ$ to 2.2 ab, which is comparable to semi-leptonic top. As described in Section~\ref{sec:signal}, these estimates for the background cross sections will also be valid at the High Luminosity LHC. 

At 33 TeV (100 TeV), $t\bar t Z$ is the dominant background, but it is still small because of the very hard missing energy cuts. The total cross section is 5 pb (46 pb). Applying just the 2 TeV (4 TeV) missing energy cut reduces the effective cross section to 91 ab (130 ab). Applying the requirement of both tops having $p_T > 500$, as well as branching ratios and $b$ and top tags reduces the effective cross section to 0.78 ab (1.4 ab). Thus we see that at higher energy, it becomes easier to separate signal from background, and stops become easier to discover as long as a collider has enough energy and luminosity to produce them. The effective background cross sections for all the different colliders are summarized in Table~\ref{tab:bg}.

\section{Estimate of Reach}

\begin{figure}
\centerline{\includegraphics[width=.5\textwidth]{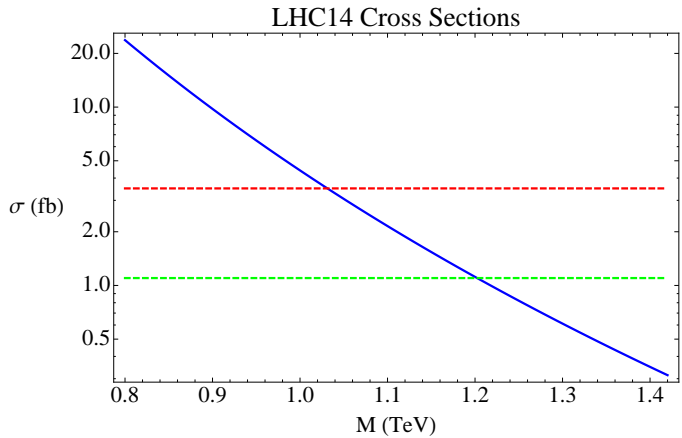}}
\caption{Tree level stop pair production cross section at the LHC14 as a function of the mass of a stop (solid blue). The horizontal dashed lines are the estimated 5$\sigma$ discovery reach with 300 fb$^{-1}$ (top red) and 3 ab$^{-1}$ (bottom green). }
\label{fig:LHC14}
\end{figure}

In this section we estimate the reach in terms of stop pair production cross section and stop mass. Our results are summarized in Table~\ref{tab:results}. We can estimate the $\sigma$-significance as number of signal events divided by the square root of the number of background events. This can be rewritten as a discovery of $N_\sigma$ being achieved with the following signal cross section
\begin{equation}
\sigma_{s} = \frac{N_\sigma}{\varepsilon_s}\sqrt{\frac{\varepsilon_b \sigma_b}{L}}
\label{eq:nsigma}
\end{equation}
where $\varepsilon_s$ is the signal efficiency computed in Section~\ref{sec:signal}, $\varepsilon_b \sigma_b$ is the effective background cross section computed in Section~\ref{sec:bg}, and $L$ is the integrated luminosity of the collider run. For the LHC with 300 fb$^{-1}$ (3 ab$^{-1}$), this corresponds to a $5\sigma$ discovery of stops with pair production cross sections of 3.5 fb (1.1 fb). We then use the tree level cross section computed by Madgraph 5~\cite{Alwall:2011uj} to translate these cross sections to mass reaches of 1.0 TeV (1.2 TeV). Figure~\ref{fig:LHC14} shows the stop cross section at $\sqrt{s}=14$ TeV as a function of mass as well as the cross sections where a 5$\sigma$ discovery is possible. 

We can repeat this analysis for a potential 33 (100) TeV collider with the uncertainties on all estimates becoming larger the further we get from current LHC running conditions. For these colliders, we are now in a regime where we can set cuts such that the expected number of background events is $O(1)$, so we need $O(5)$ events for a 5$\sigma$ discovery. In this regime, Eq.~(\ref{eq:nsigma}) is not strictly correct, but will suffice as a reasonable approximation here. Taking the results from Table~\ref{tab:bg}, we find that we can discover stops with pair production cross section of 91 ab (200 ab), which we can translate to a mass reach of 3.0 TeV (5.7 TeV).

This is of course just a simple estimate of the reach, and there many things that could be done to make it more precise including implementing top decays and hadronization as well as a realistic detector simulation. One can also consider more sophisticated cuts which vary for the different collider scenarios. It would also be interesting to see what the top tagging efficiency and fake rates look like at even higher top momenta. We leave these and the other issues not considered for future study.

\section*{Acknowledgments}
DS thanks Kaustubh Agashe for helpful conversations and comments on the draft, and also thanks Yang Bai, Kirill Melnikov, and Maxim Perelstein for useful conversations, and all the Snowmass organizers and participants for all their hard work on this community study. DS is supported in part by the NSF under grant PHY-0910467, gratefully acknowledges support from the Maryland Center for Fundamental Physics, and is appreciative of support and hospitality from the Kavli Institute for Theoretical Physics where part of this work was completed.


\bibliography{lit}
\bibliographystyle{JHEP}

\end{document}